\documentclass[journal]{IEEEtran}
\usepackage{tensor}
\usepackage{graphicx}
\usepackage{tabularx}
\usepackage[cmex10]{amsmath}
\usepackage{amsthm}
\usepackage{amssymb}
\usepackage{bm}
\usepackage{algorithm}
\usepackage{algorithmic}
\usepackage{setspace}
\usepackage{stfloats}
\usepackage{enumerate}
\usepackage{cases}
\usepackage{multirow}
\usepackage{mathrsfs}
\usepackage{array}
\usepackage{color}
\usepackage{verbatim}
\usepackage{extpfeil} 
\usepackage{booktabs}
\usepackage{graphicx}
\usepackage{tabularx}
\usepackage{epstopdf}
\usepackage{textcomp}
\usepackage{epsfig,epsf,color,balance,cite}
\usepackage{amsmath}
\usepackage{epsfig,epsf,color,balance,cite}
\usepackage{textcomp}
\usepackage{url}
\usepackage{tikz}  
\usepackage[colorlinks,citecolor=blue,linkcolor=blue,urlcolor=blue]{hyperref}

\usepackage{booktabs,tabularx,makecell}
\renewcommand{\arraystretch}{1.05}

\def\BibTeX{{\rm B\kern-.05em{\sc i\kern-.025em b}\kern-.08em
		T\kern-.1667em\lower.7ex\hbox{E}\kern-.125emX}}
\begin{document}

\title{DOA Estimation for Low-Altitude Networks: HAD Architectures, Methods, and Challenges}

\author{Ye Tian, Tuo Wu, Jintao Wu,  He Xu, Yuanjun Shen,  Xianfu Lei,  and Kin-Fai Tong,  \emph{Fellow, IEEE}

\vspace{-5mm} 
\thanks{(\textit{Corresponding author: Tuo Wu.})}

\thanks{This work was partially supported by the Natural Science Foundation of China under Grant 62571286, the Zhejiang Provincial Natural Science Foundation of China under Grant LQN26F010011, and the Natural Science Foundation of Ningbo Municipality under Grant 2024J232. The work of T. Wu was funded by Hong Kong Research Grants Council under the Area of Excellence Scheme under Grant AoE/E-101/23-N. The work of K. F. Tong was funded by the Hong Kong Metropolitan University, Staff Research Startup Fund: FRSF/2024/03.

Y. Tian and J. Wu are with the School of Artificial Intelligence, Ningbo University, Ningbo, China (e-mail: $\rm \{tianye1, 2411100177\}@nbu.edu.cn$).
T. Wu is with the Department of Electronic Engineering, City University of Hong Kong, Hong Kong (e-mail: $\rm tuo.wu2@cityu.edu.hk$).
H. Xu is with the School of Cyber Science and Engineering, Ningbo University of Technology, Ningbo, China (e-mail: $\rm xuhebest@sina.com$). Y. Shen is with National Key Laboratory of Radar Detection and Sensing, Department of Electronic Engineering, Xidian University, Xi’an, 710071, China. (e-mail: $\rm yuanjun.shen@xidian.edu.cn$). X. Lei is with the School of Information Science and Technology, Southwest Jiaotong University, Chengdu 610031, China (e-mail: $\rm xflei@swjtu.edu.cn$). K. F. Tong is with the School of Science and Technology, Hong Kong Metropolitan University, Hong Kong SAR, China. (e-mail: $\rm ktong@hkmu.edu.hk$).} 

}

\maketitle

\begin{abstract}
	With the rapid expansion of low-altitude economy (LAE) services and the growing demand for integrated sensing and communication (ISAC) in air-ground networks, reliable direction-of-arrival (DOA) estimation has become essential for both directional communication and sensing functions. DOA underpins beam alignment, spatial-reuse scheduling, and ISAC-critical tasks such as airspace situational awareness and multi-target monitoring. Hybrid analog-digital (HAD) architectures have emerged as a practical solution for large-aperture directional operation under stringent radio frequency (RF), analog-to-digital converter (ADC), and size, weight, and power (SWaP) constraints. However, HAD compresses antenna-domain observations through analog combining, fundamentally reshaping the measurement model and introducing new algorithmic and system-level challenges for DOA estimation. This article first reviews the principles and representative architectures of HAD, highlighting their advantages for scalable beam-centric and ISAC-oriented operation in LAE scenarios. We then provide a structured overview of HAD-enabled DOA estimation methodologies, including spatial covariance matrix (SCM) reconstruction, multi-combiner scan-based acquisition, and pilot-aided estimation, along with key design tradeoffs. Finally, we discuss open challenges and outline reliability-driven research directions toward robust, deployable HAD-enabled DOA solutions for practical ISAC-enabled low-altitude environments.
\end{abstract}


\section{Introduction}
Low-altitude operations are rapidly emerging as a strategic growth driver, spanning logistics, inspection, emergency response, and advanced air mobility (AAM), thereby accelerating the low-altitude economy (LAE) ecosystem \cite{r1,r2}. These diverse missions demand reliable command-and-control links, high-throughput payload transmission along predefined routes, and stable connectivity during close-range maneuvering under frequent blockage from buildings and terrain; emergency response further relies on rapid network deployment in uncertain environments, supported by resilient airborne relays. As a result, low-altitude networks face stringent constraints, including severe size, weight, and power (SWaP) limitations on airborne platforms and highly dynamic propagation conditions. Moreover, low-altitude systems prioritize not only instantaneous throughput and coverage but also service continuity and operational safety. Consequently, the ability to rapidly and reliably acquire angular-domain information has become foundational for the co-design of low-altitude communications and sensing, and is increasingly recognized as a shared physical-layer primitive for ISAC in low-altitude networks.

At millimeter-wave and sub-terahertz bands, abundant bandwidth comes at the cost of higher path loss and stronger directionality, making accurate beam alignment a prerequisite rather than an optional optimization \cite{r3}. Meanwhile, integrated sensing and communication (ISAC) functions such as airspace situational awareness, multi-target monitoring, and interference-source localization also depend on reliable angular information from array measurements \cite{r4}, \cite{r8}. Therefore, direction-of-arrival (DOA) estimation, as a fundamental capability for angular awareness and positioning, becomes a core enabler for low-altitude ISAC systems: accurate DOA supports not only beam alignment and spatial-reuse scheduling on the communication side but also target localization, trajectory tracking, and interference-source identification on the sensing side \cite{r5}.
\begin{figure*}[htbp]
	\centering 
	\includegraphics[width=1\textwidth]{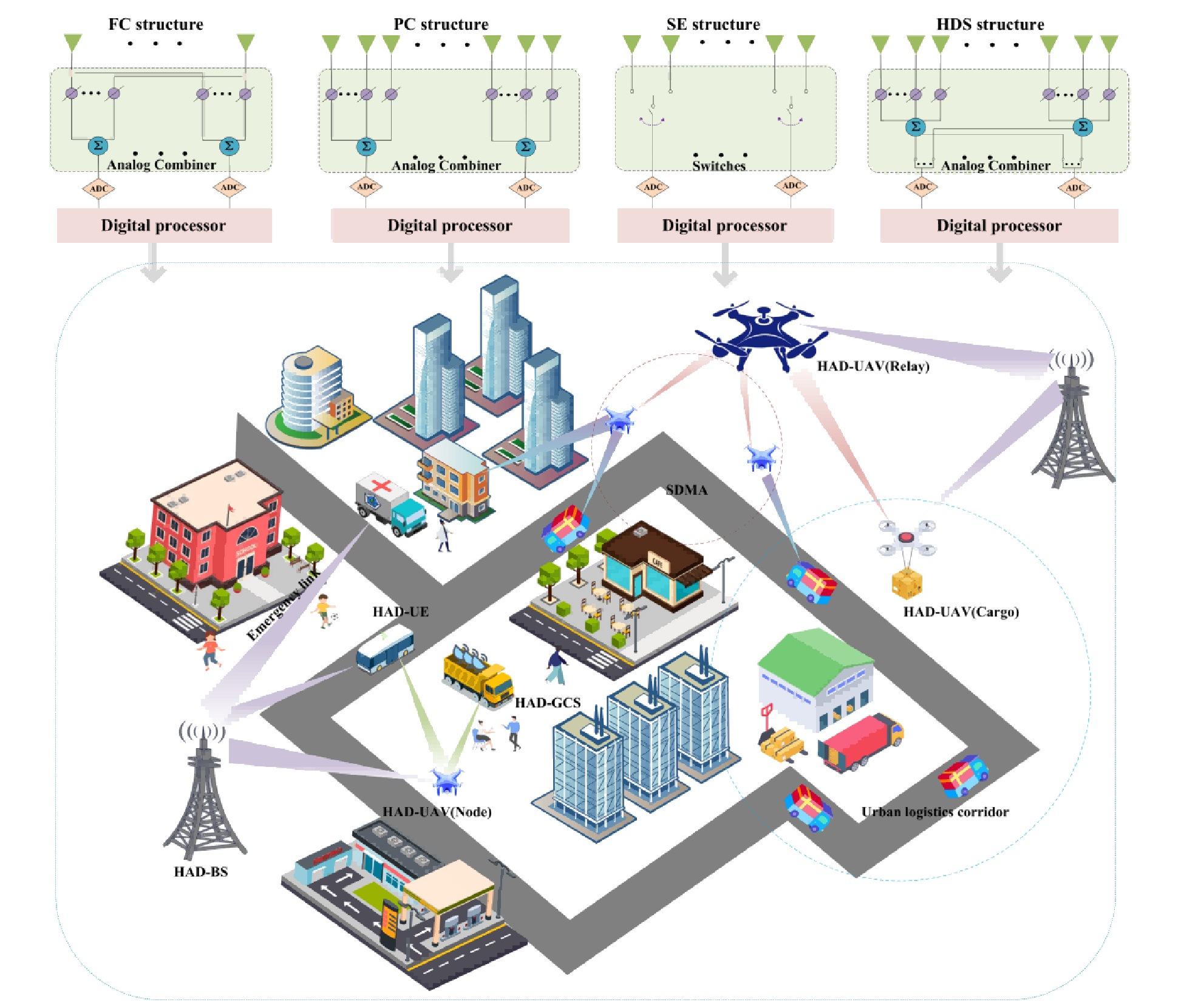}
	\caption{Representative LAE scenarios enabled by HAD nodes.} 
	\label{MX} 
\end{figure*}

However, the operational requirements of LAE can make fully digital (FD) architectures impractical. In low-altitude corridors, frequent beam updates are necessary to maintain directional links under mobility and blockage; in swarm scenarios, multiple airborne nodes require angular discrimination to enable spatial reuse and suppress co-channel interference. Both cases demand sustained high-rate acquisition and processing of high-dimensional multi-antenna data, leading to substantial RF/ADC power consumption and baseband computation burden. This system-level scalability bottleneck makes FD arrays prohibitively costly precisely in the LAE regimes where directional links and reliable angular awareness are most needed.

Hybrid analog-digital (HAD) architectures thus serve as a practical solution for LAE systems. By introducing an analog combining network, HAD significantly reduces the required RF chains and high-speed ADCs, lowering power consumption and baseband throughput while leveraging large-aperture arrays for directional gain. However, this dimensionality reduction alters the observation model and prevents direct reuse of conventional FD DOA estimation methods \cite{r7}. Consequently, DOA estimation must be redesigned to account for HAD measurements. In this article, we first introduce the operating principles and representative architectures of HAD, along with their deployment roles in ISAC-oriented LAE networks. We then review DOA estimation methods for HAD systems. Finally, we discuss open challenges and outline reliability-driven research directions toward deployable ISAC-enabled low-altitude solutions.
\section{Why HAD Architectures for Low-Altitude?}
This section addresses why HAD architectures are a compelling choice for low-altitude airspace operations and DOA estimation. We first introduce the operating principle of HAD and the associated architectural options, then discuss their deployment roles and system-level suitability for dynamic 3D deployments, as illustrated in Fig.~\ref{MX}.
\subsection{What Is HAD?}
Unlike FD receivers that place an independent RF chain and high-resolution ADC behind each antenna element, HAD architectures split array processing into two stages. First, an analog combining network in the RF domain performs weighted aggregation and dimensionality reduction, mapping high-dimensional antenna observations into low-dimensional outputs across a small number of RF chains. Then, baseband digital processing is applied to these combined signals for beamforming, detection, and estimation. This layered design substantially reduces the required RF chains and ADCs, thereby alleviating power consumption, cost, and data-interface throughput burdens, making large-aperture directional gain realizable under stringent SWaP constraints in low-altitude deployments. In practice, HAD systems switch among multiple combiner configurations to obtain sufficient angular observability for DOA estimation and beam management. Consequently, both algorithm design and system performance are tightly coupled to the analog-network realization. Below, we introduce representative HAD architectures and discuss their tradeoffs in LAE deployments.
\subsection{Representative HAD Architectures}

	\begin{table}[!t]
		\caption{Representative HAD Architectures (FC/PC/SE/HDS) and Key Characteristics}
		\label{tab:had_tradeoffs}
		\centering
		\footnotesize 
		\setlength{\tabcolsep}{5pt}  
		\renewcommand{\arraystretch}{1.6}
		\begin{tabular}{p{0.15\textwidth} p{0.06\textwidth} p{0.06\textwidth} p{0.06\textwidth} p{0.06\textwidth}}
		\hline
		\textbf{Characteristics} & \textbf{FC} & \textbf{PC} & \textbf{SE} & \textbf{HDS} \\
		\hline
		Topology flexibility  & -- & -- & Medium & High \\
			\hline
	Control method & PS network & Sub-PS network& Switch network & Switch + PS  \\
			\hline
		Power consumption & High & Medium & Low & Medium--high \\
			\hline
		Combining flexibility &  High & Limited & Limited & Very high \\
		\hline
		Best fit (LAE) & BS /\newline sensing & BS/airborne & UE/GCS / access & BS/airborne/ \newline tracking \\
		\hline
	\end{tabular}
	\end{table}

\subsubsection{Fully Connected Architecture}
In a fully connected HAD (FC-HAD) architecture, each RF chain connects to the entire antenna array through a dense phase-shifter (PS) network. Because every chain accesses the full aperture, FC-HAD offers high array gain and fine-grained beam control, enabling flexible beam shaping, interference nulling, and multi-beam operation. These capabilities make FC-HAD attractive when high angular selectivity is needed for dense 3D airspace, such as ground base stations serving low-altitude corridors or sensing nodes supporting multi-target monitoring.

However, FC-HAD incurs a high PS count and dense RF routing, increasing insertion loss and calibration burden. These drawbacks translate into higher cost and power consumption, particularly in wideband and high-frequency operation, where frequency-dependent impairments can further reduce beamforming efficiency. Consequently, FC-HAD is most suitable for infrastructure-side deployments where SWaP constraints are relaxed and peak spatial selectivity justifies the hardware complexity.

\subsubsection{Partially Connected Architecture}
In a partially connected HAD (PC-HAD) architecture, each RF chain connects only to a dedicated subarray, substantially simplifying the PS network compared with FC-HAD. This structure reduces routing complexity and improves power efficiency \cite{r9}. As a result, PC-HAD is well suited for airborne terminals and compact platforms in LAE, where energy budget, thermal dissipation, and form-factor constraints dominate.

Since each RF chain observes only a portion of the full aperture, combining freedom is reduced, limiting multi-beam capability and compromising angular discrimination relative to FC-HAD under the same array size and RF-chain count—particularly when conventional estimators are applied without exploiting additional diversity, training, or reconstruction. In practice, PC-HAD enables directional access and tracking with bounded power while partially recovering angular fidelity through multi-combiner designs and pilot-aided updates. This makes PC-HAD a common choice for large-scale base station (BS) deployments and airborne platforms that prioritize robust tracking and fast re-acquisition over maximum resolution.

\subsubsection{Switches-Based Architecture}
Switches-based HAD (SE-HAD) architecture replaces PS networks with RF switches, allowing each RF chain to connect to a selected subset of antennas and rapidly update this selection. Switch networks are typically low-loss and fast to reconfigure, making SE-HAD suitable for coarse spatial probing, event-triggered sensing, and initial-access procedures in low-altitude networks \cite{r10}. From an LAE viewpoint, SE-HAD is appealing for SWaP-limited terminals and portable user equipment (UE)/ground control station (GCS) devices, where low-duty-cycle operation and fast bring-up are more valuable than fine-grained beam synthesis. The main limitation is reduced instantaneous aperture access and limited combining freedom per slot, which can lower array gain and estimation accuracy—especially in multipath-rich urban canyons. Therefore, SE-HAD is best viewed as a low-overhead front-end for directional probing and quick recovery, rather than a stand-alone solution for high-resolution DOA estimation.

\subsubsection{Hybrid Dynamic Subarray Architecture}
The hybrid dynamic subarray (HDS) architecture bridges PC-HAD and FC-HAD. In HDS, the receive array is partitioned into multiple subarrays \cite{r11}, \cite{r12}, and a switch network is embedded between the RF chains and subarrays, enabling each RF chain to connect to any subarray through programmable switches. With phase-controlled analog combining on active connections, the resulting combiner supports phase-controllable hybrid processing. The defining feature of HDS is its adjustable switch-closure ratio: a higher ratio increases effective aperture utilization and preserves more spatial information, whereas a lower ratio reduces RF-network complexity and power consumption. By tuning the closure ratio, HDS can smoothly transition between PC-like and FC-like operating modes.

From an LAE deployment perspective, HDS tunability enables switching between competing requirements: in low-altitude corridor operations and urban-canyon flight, frequent blockage demands rapid re-acquisition, favoring a low-closure, PC-like configuration; in contrast, swarm access or interference-limited multi-target operations require higher angular resolution, making a high-closure, FC-like configuration more suitable. In practice, HDS is best suited to infrastructure-side nodes (e.g., corridor base stations and fixed sensing sites) and capable airborne relays. Overall, HDS represents a promising architecture for DOA estimation and beam management in dynamic low-altitude environments.
\subsection{HAD-Enabled LAE Applications}
Fig.~\ref{MX} illustrates a representative LAE scene where HAD nodes coexist across infrastructure, airborne platforms, and terminal-side control units, while Table~\ref{tab:fd_had_lowalt_overview} summarizes capability-application-challenge tradeoffs against FD counterparts. Below, we map the roles in Fig.~\ref{MX} to the HAD categories in Table~\ref{tab:fd_had_lowalt_overview} and highlight how DOA estimation supports practical LAE operations.

\subsubsection{HAD-BS}
As shown in Fig.~\ref{MX}, HAD-BS provides scalable directional coverage and high angular discrimination for corridor operation and dense swarm access, approaching FD angular resolution with far fewer RF chains. DOA enables multi-user angular separation for spatial division multiple access (SDMA), interference-aware beamforming/nulling, and sectorized scheduling. The dominant challenges are combiner identifiability and robustness to hardware non-idealities; consequently, structured covariance/subspace processing with multi-combiner fusion and calibration-aware designs are critical.

\subsubsection{HAD-Airborne}
The HAD-Airborne nodes in Fig.~\ref{MX} emphasize mobility-centric DOA: fast acquisition, tracking-friendly updates, and rapid re-acquisition under blockage. The primary benefit is enabling directional access and tracking with a lightweight RF front-end under strict payload, power, and thermal constraints. Unlike the BS case, airborne-side DOA prioritizes low overhead and stable maintenance; therefore, scan-based acquisition and coarse-to-fine refinement are commonly adopted. The main challenges include short coherence time, vibration-induced impairments, and wideband effects under fast motion.

\subsubsection{HAD-UE/GCS}
The HAD-UE/GCS in Fig.~\ref{MX} represents terminal-side control, most beneficial when compact multi-antenna panels are available (e.g., portable ground control stations). UE/GCS-side DOA is typically not aimed at ultra-high resolution due to limited aperture; instead, it supports fast beam alignment, directional interference rejection, and rapid re-acquisition in cluttered environments, often by providing beam/alignment feedback to the network. Accordingly, UE/GCS implementations favor simple codebook-based scanning and lightweight refinement/tracking. Key challenges include limited identifiability under multipath and interference, and wideband beam-squint at higher carrier frequencies, calling for low-overhead yet robust beam-management procedures.

Taken together, Fig.~\ref{MX} and Table~\ref{tab:fd_had_lowalt_overview} suggest a function-driven view: HAD-BS targets high-resolution angular separation, HAD-Airborne prioritizes fast acquisition and tracking stability, and HAD-UE/GCS emphasizes low-overhead directional connectivity. This role-specific mapping explains why different HAD-DOA algorithm families are preferred for different platforms.
 
 \begin{table*}[t]
 	\centering
 	\caption{Unified Overview: FD vs. HAD Architectures in LAE with Capabilities, Applications, and Challenges}
 	\label{tab:fd_had_lowalt_overview}
 	\renewcommand{\arraystretch}{1.12}
 	\setlength{\tabcolsep}{5.5pt}
 	\begin{tabularx}{\textwidth}{p{2.1cm} p{3.0cm} X X X}
 		\hline
 		\textbf{Category} & \textbf{Dimension} & \textbf{HAD-BS vs FD-BS} & \textbf{HAD-Airborne vs FD-Airborne} & \textbf{HAD-UE/GCS vs FD-UE/GCS} \\
 		\hline
 		
 		\multirow{3}{*}{\textbf{Capabilities}}
 	& Beamforming / DOA
 	& Near-FD resolution; fewer RF chains
 	& Directional access; tracking-friendly
 	& Multi-panel gain; alignment support \\
 		
 	& Complexity / Power
 	& RF/ADC power cut; lower baseband rate
 	& Endurance gain; SWaP-aware
 	& Low cost/power; portable-friendly \\
 	& Robustness
 	& Multi-combiner fusion; structured stats
 	& Fast update; stable tracking
 	& Interference rejection; re-acquisition \\
 		\hline
 		
 		\multirow{4}{*}{\textbf{Applications}}
 	& Swarm access / spatial reuse
 	& Angular separation enables SDMA, corridor scheduling, and interference nulling.
 	& Supports formation links and local close-angle maintenance.
 	& Mainly supports alignment and feedback rather than scheduling. \\
 	
 	& Emergency / rapid deployment
 	& Wide-area coverage with fast discovery and alignment.
 	& Airborne relay/BS with feasible arrays for coverage extension and tracking.
 	& Compact directional modules for quick field setup. \\
 	
 	& Inspection / sensing-assisted comm
 	& Beam training plus angular awareness for tracking and monitoring.
 	& Shared comm/sensing front-end for route-following and obstacle-aware links.
 	& Angle-aware link monitoring; sensing limited by aperture. \\
 	
 	& Urban logistics corridor
 	& Low-overhead frequent updates under dense reuse.
 	& Route-following updates with coarse-to-fine refinement.
 	& Reliable control/telemetry and fast re-acquisition. \\
 		\hline
 		
 		\multirow{4}{*}{\textbf{Challenges}}
 		& Combiner design / identifiability
 		& Ambiguity; ill-conditioning; overhead--resolution
 		& Short coherence; motion-induced switching errors
 		& Simple combiners; small-aperture multipath  \\
 		
 		& Covariance / subspace reconstruction
 	& Toeplitz/low-rank; conditioning control
 	& Lightweight; partial subspace
 	& Reconstruction hard; priors needed \\
 		
 		& Wideband / near-field effects
 		& Beam squint; near-field (large aperture)
 		& Wideband+motion; near inspection
 		& Wideband; limited near-field modeling \\
 		
 		& System co-design
 		& Joint training, scheduling, and DOA updates to realize spatial-reuse gains.
 		& Integrate DOA with flight dynamics and link adaptation; avoid frequent re-training.
 		& Lightweight protocols with minimal overhead; reliability over peak capacity. \\
 		\hline
 	\end{tabularx}
 \end{table*}

\section{HAD-Enabled DOA Estimation}
The inherent compression of antenna-domain observations in HAD receivers fundamentally alters the measurement model, making conventional DOA estimators designed for FD arrays no longer directly applicable. This section surveys representative HAD-enabled DOA estimation approaches, organized into three complementary families: spatial covariance matrix (SCM) reconstruction, beamforming-based scanning with coarse-to-fine refinement, and pilot-aided schemes that achieve high-resolution DOA under limited RF chains.

\subsection{Spatial Covariance Matrix Reconstruction}
In HAD architectures, the analog combining network compresses high-dimensional antenna-domain observations into a small number of RF-chain outputs. This compression alters the effective observation model, making it difficult to directly reuse the conventional FD pipeline that first forms a full-dimensional SCM before performing subspace decomposition. If subspace-based algorithms are applied directly to compressed hybrid measurements, their performance degrades noticeably at low SNR or when few RF chains are available. In contrast, once an equivalent SCM is reconstructed and fed into a subspace method, DOA estimation accuracy can approach the FD baseline over a wide SNR range and becomes substantially less sensitive to the number of RF chains. Consequently, in LAE scenarios where high-resolution angular separation and spatial multiplexing are crucial, SCM reconstruction has emerged as a key technique to translate HAD hardware scalability into practical angular discrimination capability.

One common approach is element-/entry-wise reconstruction. The main idea is to view the low-dimensional covariance observed in the hybrid domain as the projection of an unknown full-dimensional covariance through the analog combiner. During training, the receiver switches among multiple complementary analog combiners across time slots to collect sufficient independent observations, after which a linear solver recovers the unknown entries. This approach is attractive because it directly targets the spatial correlation structure, beneficial for resolving closely spaced angles and enabling multi-user angular-domain decoupling. However, its effectiveness depends strongly on combiner-set identifiability; if different combiners are highly correlated or the resulting linear system is ill-conditioned, reconstruction may amplify noise and hardware mismatches, leading to unstable solutions or ambiguities.

Another widely used route is beamforming-based reconstruction, which recovers covariance blocks rather than treating every SCM element independently. The hybrid receiver switches among predefined analog combiners so that analog beams sequentially point to different candidate directions; for each configuration, it computes correlation statistics to produce low-dimensional measurements that vary with beam direction \cite{r13}. The receiver then reconstructs sub-SCMs via linear inversion and assembles them into the full SCM for subspace-based DOA estimation (e.g., MUSIC), reusing the training procedure without additional hardware complexity.

When a uniform linear array (ULA) is employed, both approaches can be simplified to reduce computational cost. Owing to the shift-invariant structure of a ULA, the SCM has constant entries along each diagonal, meaning the number of degrees of freedom (DoF) to be recovered is far smaller than the matrix dimension. In entry-wise reconstruction, one can recover only a small set of non-redundant lag-correlation parameters, which are then filled according to the structural rule to form Toeplitz sub-blocks. In beamforming-based reconstruction, each sub-SCM block can similarly be constrained to a structured form, converting the recovery task into the estimation of only a few structural parameters. This reduces training requirements and improves numerical stability.

\subsection{Beamforming-Based DOA Estimation}
In low-altitude scenarios, DOA estimation on the airborne side typically prioritizes rapid capture, continuous tracking, and fast re-acquisition after blockage. On the UE/GCS side, constraints on size, power consumption, and implementation complexity make lightweight angular-domain scanning the preferred option for beam alignment and link maintenance. Rather than reconstructing the full-dimensional spatial covariance at every update, a more common strategy is to first obtain coarse DOA information via low-overhead scanning and then perform local refinement within a candidate angular region, achieving a controllable trade-off among training overhead, real-time responsiveness, and reliability.

During the coarse-estimation stage, beamforming-based methods are particularly suitable. Because HAD architectures observe only low-dimensional outputs after analog combining, the receiver switches among different analog combining matrices across training/probing time slots to form analog beams pointing to different angular regions. With a few time slots, the receiver can rapidly scan the spatial domain, output candidate beams or spatial sectors, and convert the most reliable candidates into coarse DOA estimates, providing initialization for subsequent refinement and tracking.

Regarding combiner design, two main approaches exist. The first is candidate-angle-driven construction, where analog combining matrices are generated from array steering vectors, and the angle with the strongest pilot-correlated response is selected as a coarse DOA estimate. The second is DFT-based multi-combiner design, which constructs approximately orthogonal combiners with near-uniform angular coverage \cite{r14}. Wide beams first screen candidate sectors; subsequently, narrow beams perform local fine sweeping. This two-stage procedure transforms a high-cost global search into a manageable process, matching low-altitude requirements for timeliness and rapid re-acquisition.

In practice, combiner-set design must balance scanning overhead and angular resolution. A robust approach is to adapt beam density to scenario dynamics: denser, narrower beams in crowded angular domains (e.g., swarms or strong interference) and sparser coverage when rapid search is needed.

\subsection{Pilot-Aided DOA Estimation}
Pilot-aided DOA estimation mitigates performance degradation from observation compression by exploiting a time-division multiplexed observation mechanism \cite{r15}. Assuming angular parameters remain approximately constant over a short time window, the transmitter sends multiple independent pilot sequences while the receiver applies different analog combining configurations across measurements. The resulting low-dimensional compressed measurements are aggregated to form an equivalent high-dimensional virtual array observation, whose dimension can be adaptively expanded with the number of pilot uses, improving estimation accuracy and robustness without increasing RF chains.

To ensure sufficient angular-domain coverage, a large-scale PS network generates randomly sampled phase coefficients so that each pilot measurement rapidly ``illuminates'' the angular domain, avoiding blind spots from fixed beams. By fusing diversified observations with virtual-array aggregation, pilot-aided HAD processing enables comprehensive angular information acquisition under hardware constraints, supporting high-precision positioning and sensing in LAE.

\section{Open Challenges and Future Directions}
Despite the promising potential of HAD arrays for scalable DOA acquisition in LAE networks, HAD-enabled DOA estimation and its system-level deployment remain in an early stage. Many fundamental issues are open; in particular, the coexistence of high mobility, intermittent blockage, dense multi-user interference, and compressed hybrid measurements makes low-altitude DOA estimation more challenging than conventional ground or static scenarios. This section discusses key open challenges and outlines future research directions.

\subsubsection{Reliability-Centric DOA Objectives}
Most existing studies focus on snapshot-level accuracy metrics (e.g., RMSE or peak resolution). However, LAE systems are often dominated by service continuity and operational safety. Moderate DOA errors are tolerable as long as beams remain stable, whereas brief loss-of-lock events or prolonged re-acquisition after blockage can severely undermine link reliability. A key challenge is that current evaluation criteria do not fully capture LAE operational objectives. Future research should move toward reliability- and control-aware objectives, such as update stability, loss-of-lock probability, and re-acquisition latency, and provide well-calibrated confidence measures so that beam management can selectively trigger refinement only when necessary.

\subsubsection{Blockage-Driven Dominant-Path Switching}
Urban canyons and LAE corridors frequently exhibit abrupt transitions between LoS components and reflected dominant paths. In such environments, DOA does not evolve smoothly but can undergo sudden jumps due to blockage and path switching. A central challenge is developing HAD-DOA pipelines that detect when the current estimate becomes unreliable, rapidly re-acquire the new dominant direction with bounded sensing and computational overhead, and provide well-calibrated confidence measures exploitable by the beam-management layer. Rather than treating the problem as purely continuous tracking, future research should integrate change-point detection with confidence-aware decision logic, triggering rapid re-acquisition only when warranted.

\subsubsection{Angular-Domain Multi-User Coupling}
LAE corridors and airborne swarms often operate in interference-dense conditions with closely spaced angular signatures. In this regime, DOA estimates are no longer independent across users: interference leakage, competing beams, and shared training resources can couple estimation processes and lead to correlated failures, such as joint mis-separation of an entire angular cluster. This challenges conventional per-link DOA processing. A promising direction is networked angular inference, where angle-domain clustering and sensing-training allocation are jointly designed to prioritize the most ambiguous angular regions, translating DOA information more directly into spatial reuse and interference-management gains.

\subsubsection{Heterogeneous-Node Cooperative Inference}
LAE deployments naturally comprise heterogeneous roles (BS, airborne relays, cargo platforms, UE/GCS), yet most DOA estimators are designed for a single receiver in isolation. A promising direction is lightweight cooperative DOA, where nodes exchange compact angular features (e.g., candidate directions or confidence measures) rather than raw I/Q samples, enabling robust fusion in NLoS-dominated areas while keeping overhead minimal.

%Hybrid arrays introduce structured distortions and time-varying non-idealities that are difficult to model perfectly, and low-altitude environments exhibit strong domain shifts across locations, weather, and traffic patterns. While learning-based methods are promising for robustness, practical deployment is limited by generalization, uncertainty calibration, and the lack of standardized data collection pipelines tailored to blockage events, mobility patterns, and interference dynamics.
%
%\subsubsection{Scenario-faithful benchmarking and reproducibility}
%Low-altitude channels depend strongly on urban geometry, trajectories, blockage statistics, and network density. Without scenario-faithful and reproducible benchmarks, results can be overly sensitive to simulation choices. This limits the community’s ability to draw reliable conclusions about which HAD-DOA pipelines are truly deployable, and which improvements are artifacts of favorable assumptions.
%Overall, advancing HAD-enabled DOA for the low-altitude economy requires shifting from isolated estimator optimization toward reliability- and system-driven co-design. The most impactful directions include confidence-aware re-acquisition, networked multi-user angular inference, cooperative operation across heterogeneous nodes, model-augmented learning for robustness, and reproducible benchmarks that reflect low-altitude operational realities.
%

\section{Case Studies}
We present two case studies highlighting both the limitations of HAD architectures and the performance gains from SCM reconstruction. A ULA with half-wavelength spacing is simulated with two far-field sources at $10^\circ$ and $60^\circ$, $N = 1000$ snapshots, and 500 Monte Carlo trials, with RMSE as the performance metric. Three methods are compared: FD-MUSIC, which applies the classical MUSIC algorithm under the FD architecture; HAD-MUSIC, which naively applies MUSIC directly to compressed HAD measurements; and SCM-MUSIC, which first reconstructs an equivalent full-dimensional SCM and then performs MUSIC \cite{r13}.

\begin{figure}[t]
	\centering 
	\includegraphics[width=0.5\textwidth]{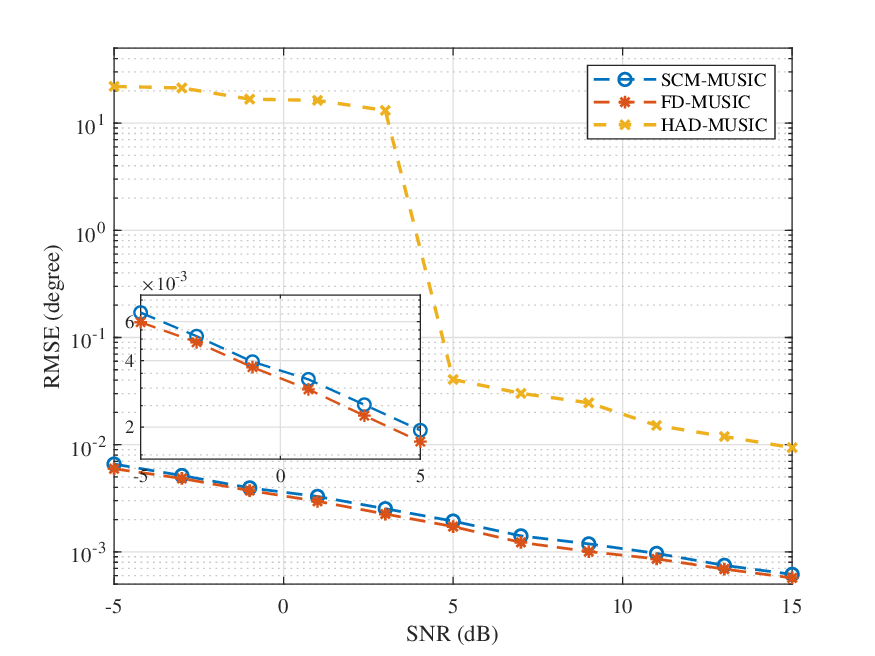}
	\caption{RMSE of DOA estimation versus SNR, with $M=64$.} 
	\label{RMSE} 
\end{figure}

Fig.~\ref{RMSE} depicts RMSE versus SNR with $M=64$ antennas. HAD-MUSIC deteriorates severely at low SNR, revealing that naive transplantation of conventional FD subspace processing to HAD can fail due to hybrid compression. In contrast, SCM-MUSIC closely approaches the FD-MUSIC benchmark over the entire SNR range, demonstrating that SCM reconstruction can recover near-FD angular accuracy with far fewer RF chains. This result carries a practical implication for LAE: even under the stringent RF-chain budgets typical of airborne and terminal-side nodes, SCM-based pipelines can deliver the angular precision needed for reliable beam alignment and ISAC-critical target discrimination.

\begin{figure}[t]
	\centering 
	\includegraphics[width=0.5\textwidth]{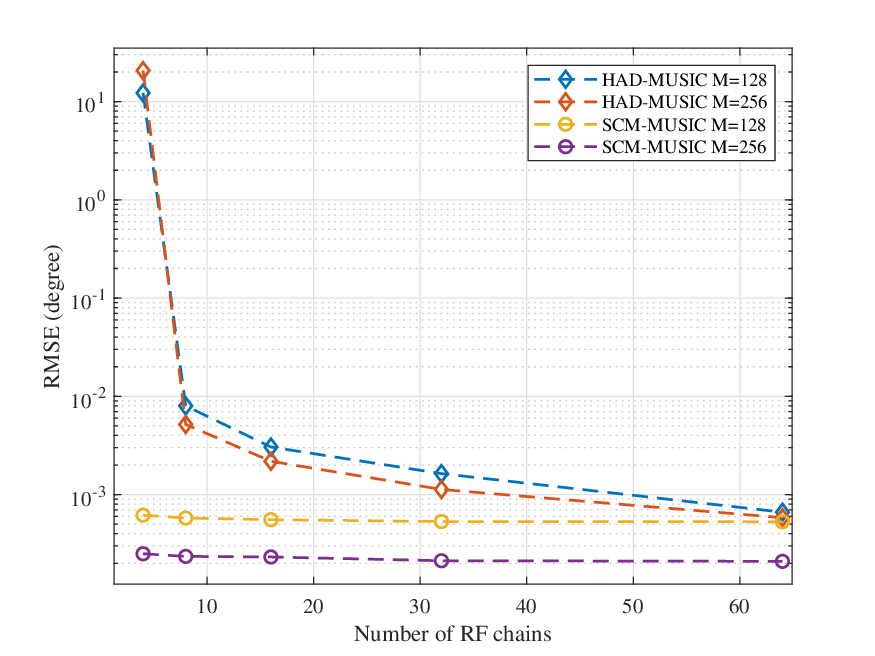}
	\caption{RMSE of DOA estimation versus $M$ and the number of RF chains.} 
	\label{L} 
\end{figure}

Fig.~\ref{L} shows how RMSE varies with RF-chain count under different antenna counts. HAD-MUSIC is highly sensitive to RF-chain count and degrades sharply when RF chains are scarce. In contrast, SCM-MUSIC maintains stable performance over a broad range of configurations. Additionally, increasing array aperture consistently improves accuracy, confirming that the primary gain originates from larger spatial aperture rather than from adding more RF chains. This insight suggests a favorable design strategy for LAE deployments: investing in a larger physical array while keeping the RF-chain count small can achieve strong DOA performance at lower SWaP cost than scaling up RF chains, which is particularly attractive for corridor base stations and airborne relays where aperture space is available but power and thermal budgets are tight.
\section{Conclusion}
LAE networks operate in highly directional, mobility-intensive, and interference-rich 3D environments, where dependable angular information is essential for fast beam acquisition/tracking, spatial reuse, and sensing/ISAC functions \cite{r6}. While FD arrays offer strong performance, their scalability is limited by RF-chain/ADC power, bandwidth, and baseband burdens, especially in wideband and high-frequency regimes. HAD architectures provide a practical alternative by delivering useful angular resolution under tightly bounded hardware resources. This article reviewed representative HAD architectures and DOA methodologies tailored to hybrid measurements, including SCM reconstruction, scan-based acquisition, and pilot-aided estimation, and mapped them to BS/Airborne/UE roles in LAE deployments. The case studies indicate that SCM reconstruction can substantially narrow the performance gap and reduce sensitivity to limited RF chains. Moving forward, deploying HAD-enabled DOA in real LAE systems will require reliability-driven design with confidence-aware re-acquisition, multi-user angular inference, lightweight cooperation, and scenario-faithful benchmarking.

\end{document}